\documentclass[]{article}
\usepackage{emulateapj}
\usepackage{psfig}
\usepackage{graphicx}

\advance \voffset by 1.00cm\relax
\def\msun{{\rm ~M}_{\odot}}

\def\mpy{{\rm ~M}_{\odot} {\rm ~yr}^{-1}}
\begin{document}

\title{ON THE CHANDRA X-RAY SOURCES IN THE GALACTIC CENTER}

\author{Krzysztof Belczynski\altaffilmark{1,2}, Ronald E. Taam\altaffilmark{1}}

\affil{
     $^{1}$ Northwestern University, Dept. of Physics \& Astronomy,
       2145 Sheridan Rd., Evanston, IL 60208\\
     $^{2}$ Lindheimer Postdoctoral Fellow\\  
     belczynski, r-taam@northwestern.edu}

\begin{abstract} 
 
Recent deep {\em Chandra} surveys of the Galactic center region have revealed the existence 
of a faint, hard X-ray source population. While the nature of this population is unknown,  
it is likely that several types of stellar objects contribute.  For sources involving
binary systems, accreting white dwarfs and accreting neutron stars with main sequence companions 
have been proposed.  Among the accreting neutron star systems, previous studies have focused on 
stellar wind-fed sources. In this paper, we point out that binary systems in which mass transfer 
occurs via Roche lobe overflow (RLOF) can also contribute to this X-ray source population. 

A binary population synthesis study of the Galactic center region has been carried out, and 
it is found that evolutionary channels for neutron star formation involving the accretion induced 
collapse of a massive ONeMg white dwarf, in addition to the core collapse of massive stars, can 
contribute to this population.  The RLOF systems would appear as transients with quiescent  
luminosities, above 2 keV, in the range from $10^{31} - 10^{32}$ ergs s$^{-1}$.  The results 
reveal that RLOF systems primarily contribute to the faint X-ray source population in the Muno 
et al. (2003) survey and wind-fed systems can contribute to the less sensitive Wang et al. 
(2002) survey. However, our results suggest that accreting neutron star systems are not likely 
to be the major contributor to the faint X-ray source population in the Galactic center. 
\end{abstract}


\section{INTRODUCTION}

Recent {\em Chandra} observations of the Galactic center region (e.g., Wang, Gotthelf \& Lang 
2002, hereinafter WGL02; Muno et al. 2003, hereinafter MM03) have led to a renewed interest in 
the understanding of compact stellar X-ray sources.  Both of these long exposure surveys have 
revealed a great number of X-ray point sources with luminosities, $L_x \lesssim 10^{35}$ 
ergs s$^{-1}$ and spectra which can be fit by an absorbed power law. 

Although the nature of these low luminosity sources is unknown, several possible interpretations 
have been proposed. In particular, Muno et al. (2004) suggest that the intermediate polar subclass 
of magnetic cataclysmic variables may dominate the faint population characterized by luminosities 
less than $10^{33}$ ergs s$^{-1}$ and a photon power law 
index ranging between 0 and 1. Among models involving accreting neutron stars (NS), Pfahl, Rappaport 
\& Podsiadlowski (2002, hereinafter PRP02) explored the hypothesis that wind-fed neutron star 
systems with intermediate and massive main sequence (MS) donors ($M \gtrsim 3 \msun$) may account 
for the majority of sources with $L_x=10^{33}-10^{35}$ ergs s$^{-1}$. Furthermore, the contributions 
from a population of pre-low mass X-ray binaries in which the neutron stars accrete from the wind 
of low mass MS companions ($\lesssim 2 \msun$) has also  been studied by Bleach (2002) and 
Willems \& Kolb (2003). However, it was recently pointed out by Popov (2003) that the neutron 
stars in this lower mass binary population may not necessarily accrete all the transferred 
material, implying that only a fraction of these types of systems would be continuous X-ray emitters. 
Among the non stellar models that have been proposed, Bykov (2003) has suggested that the faint hard 
sources result from the interaction of fast moving supernova ejecta fragments with the dense 
interstellar medium of the Galactic center region. These sources, due to their low intrinsic 
brightness, may only contribute to the deeper MM03 survey. 

The aforementioned stellar studies focused on only wind-fed NS systems, however there are a 
number of RLOF transient systems which, in quiescence, emit at X-ray luminosities very similar to 
those of the many point sources observed in the Galactic center. Specifically, transients with 
neutron stars in their quiescent state are observed to have X-ray luminosities in the range of 
$10^{32}-10^{33}$ ergs s$^{-1}$ (see Campana \& Stella 2003), while transient systems with 
black hole (BH) accretors tend to be observed at lower luminosities of $\sim 2 
\times 10^{31}$ ergs s$^{-1}$ (e.g., Tomsick et al. 2003) in their inactive state.  Although many 
of the NS transient systems are characterized by soft spectra in their quiescent state and are 
difficult to detect from the Galactic center region, some fraction have a power law component extending 
beyond 2 keV with luminosities in the range of $\sim 10^{31}-10^{32}$ ergs s$^{-1}$ (see Jonker, 
Wijnands, \& van der Klis 2004). Such systems may have been detected in the deeper survey of MM03. 
The systems with BH 
accretors are also known to emit at energies greater than 2 keV and may also contribute to the 
faint source population at this luminosity level.

X-ray transient systems with large outburst to quiescent flux ratios are known to have  either a 
NS or BH accretor and a predominantly low mass companion (classified as a low mass X-ray binary, 
LMXB).  The outburst phenomenon is generally believed to result from a thermal/viscous instability 
in an accretion disk surrounding the compact object. Only those systems with mass transfer rates 
below a critical value (e.g., Dubus et al. 1999; Menou, Perna \& Hernquist 2002) can exhibit such 
transient behavior.  In general, the donors of LMXBs are believed to be either low mass MS or 
giant-like stars.  However, observations of ultra-short period LMXBs (with orbital periods less 
than about 80 minutes), so called ultracompacts, have revealed that, at least, some of them must 
host very low mass ($0.01-0.1 \msun$) white dwarf (WD) donors (e.g., Markwardt et al. 2002, Galloway et 
al. 2002; see also Deloye \& Bildsten 2003). In a 
recent study Belczynski \& Taam (2004) showed that the population of LMXBs with low mass WD 
donors (and thus with very low mass transfer rates leading to transient behavior) may be more 
significant than previously anticipated.  Since the majority of transient sources are expected 
to be in quiescence due to the very low X-ray duty cycle, we suggest that a fraction of the low 
luminosity sources observed in the Galactic center region may reflect a population of transient 
systems in their quiescent state, while the same systems in outburst may contribute to a bright 
source population.

A binary population synthesis model {\em StarTrack} (presented in \S\,2) is used to determine the 
number of X-ray binaries (NS or BH accretor, wind-fed or RLOF) in the Galactic center region. 
The entire mass spectrum of evolved and unevolved donors is considered.  In particular, low, 
intermediate, and high-mass MS stars are investigated to account for the systems considered in previous 
studies. The results of our studies are described in \S\,3 and their implications are discussed in the 
final section.

\section{MODEL DESCRIPTION}

The {\em StarTrack} population synthesis code developed originally 
for the modeling of binaries with two compact objects (Belczynski, Kalogera \&
Bulik 2002) is used.  This code has recently undergone major revisions (Belczynski 
et al. 2004, in preparation), reflecting the new features 
associated with the improved treatments of the physical processes important for the
formation and evolution of X-ray binary systems. 
Specifically, the additions and revisions include a detailed treatment of 
tidal synchronization and circularization, individual treatment of various 
RLOF mass transfer phases, and a full numerical orbit evolution with angular 
momentum losses due to magnetic braking (MB), gravitational radiation (GR), mass transfer/loss 
and tides.  

In contrast to population synthesis studies of other groups, we also include the 
possibility that neutron star formation can occur via an accretion induced collapse 
(AIC) of a massive ONeMg white dwarf (Belczynski \& Taam 2004).  Such white dwarfs 
in binaries form from stars within the mass range of $\sim 7 - 12 \msun$ 
(Belczynski \& Taam 2004; Tauris \& van den Heuvel 2003; see also Nomoto \& 
Hashimoto 1988) which is wider than for the formation of ONeMg cores in single stars 
($\sim 8 - 10 \msun$) due to tidal mass loss (Nomoto \& Kondo 1991) and rejuvention 
(Belczynksi \& Taam 2004).
The ability to accumulate matter on the white dwarf is enhanced, paradoxically, by  
the effect of an optically thick wind from the white dwarf surface, which can 
stabilize the mass transfer in the system at high mass transfer rates 
(see Kato \& Hachisu 1994, Hachisu, Kato, \& Nomoto 1999). The accumulation 
ratio of hydrogen-rich and helium-rich matter is taken from Hachisu, Kato, \& Nomoto 
(1999) and Kato \& Hachisu (1999) respectively (see also Ivanova \& Taam 2004). For 
the direct accretion of helium 
or carbon/oxygen matter onto the ONeMg white dwarfs we make use of the work of 
Kawai, Saio, \& Nomoto (1987) in determining the evolution of the accreting 
white dwarf.  Our population synthesis study of the Galactic center should, therefore, 
be considered complementary to that of PRP02 who did not consider the AIC process
nor RLOF X-ray sources. 

The tidal effects have been calibrated using the observations of binary periods 
and eccentricities in several stellar clusters (Mathieu et al.\ 1992), recovering   
the orbital decay rate in high-mass X-ray binaries (Levine, Rappaport \& Zojcheski 2000).
The {\em StarTrack} RLOF calculation involves the detailed calculation of the mass  
transfer rates based on radius-mass exponents calculated both for the donor stars 
and their Roche lobes. 
The results of our calculations have been compared to a set of published RLOF sequences 
(e.g., Beer \& Podsiadlowski 2002) as well 
as to calculations obtained with an updated stellar evolution code (Ivanova et al. 2003).
Our approach to the mass transfer calculations allows for the possibility of (i) 
conservative versus non-conservative RLOF episodes (ii) thermally driven RLOF 
versus nuclear/MB/GR losses driven RLOF and (iii) a separation of systems as 
persistent or transient depending on whether the donor RLOF mass
transfer rate lies below the critical rate for instability to develop in the 
accretion disk (adopted from Dubus et al. 1999 or Menou et al. 2002 for different
compositions of transferred material). 
Persistent X-ray emitters remain at a X-ray luminosity level
corresponding to the secular RLOF mass transfer rate. Transient
X-ray systems remain at very low luminosities for most of their lifetime, however,  
they exhibit outbursts with luminosities that can reach the Eddington limit. 
We account for wind accretion onto compact objects following Hurley, 
Tout \& Pols (2002) with  the wind mass loss rates adopted from Hurley, Pols
\& Tout (2000) extended to include low mass and intermediate mass MS stars utilizing 
the formulae of Nieuwenhuijzen \& de Jager (1990). 
The wind velocities are assumed to be proportional to the escape velocity from the 
stellar surface with a normalization proposed by Hurley et al. (2002). 
For all wind-fed systems, we assume an orbit-averaged accretion rate. 
The flaring behavior associated with B{\em e} stars and the periastron passage 
enhanced mass accretion onto the NS are not taken into account.

A large sample of single ($10^6$) and binary ($10^6$) stars in the Galactic 
center are evolved for 10 Gyrs assuming a constant star formation rate. 
We choose parameters from the reference model in Belczynski 
et al. (2002) with a few differences. Specifically, the maximum NS mass is set equal to 
$2 \msun$, and the most recently inferred natal NS kick distribution (for NSs formed by the 
core collapse of massive stars) of Arzoumanian, Chernoff \& Cordes (2002) is incorporated. 
For cases in which NS formation occurs via the accretion induced collapse 
evolutionary channel, no kicks are applied (see Belczynski \& Taam 2004). 
However, we calculate one model in which we allow for the full kicks
imparted to NSs formed through AIC, to assess the effect of our {\em no AIC
kick} assumption.  In the primordial progenitor population, the primary masses are selected in 
the range $4-100 \msun$ (with an initial-mass-function slope of -2.7; see 
Kroupa, Tout \& Gilmore 1993), and the secondary masses are selected in the range $0.08-100 
\msun$ with a flat mass ratio (secondary divided by primary mass) distribution.
In the rapid phases of binary evolution, the accretion rate onto the NS and the BH is 
limited at the maximum Eddington limit with the remaining transferred material lost with a 
specific orbital angular momentum of the accretor, but allowing for hyper-critical accretion 
in the common envelope (CE) phases (e.g., Blondin 1986; Chevalier 1989; Brown 1995). In addition to  
the dynamically unstable RLOF mass transfer events leading to a CE phase, we also allow for 
evolution into the CE phase in the cases in which the trapping radius of the accretion flow exceeds 
the Roche lobe radius of the accretor (e.g., King \& Begelman 1999; Ivanova et al. 2003).
Finally, in comparison with our previous study, we have assumed that all systems that  
enter a CE phase with the donor in the Hertzsprung gap merge into a single star
(Ivanova \& Taam 2004; Belczynski \& Taam 2004). 

The updated version of the code has been tested and used for a study of Galactic ultra compact 
binaries (Belczynski \& Taam 2004), Galactic young black hole populations (Belczynski, Sadowski 
\& Rasio 2004) and used to reproduce the X-ray luminosity function of the 
nearby starburst galaxy NGC 1569 (Belczynski et al. 2004).

\section{RESULTS}

\subsection{Source Type and X-ray Luminosity}

In our study only accreting binaries with NS and BH primaries, which are brighter than $L_x 
= 10^{30}$ ergs s$^{-1}$ are considered.  The secondaries in these X-ray binaries may lose 
material either through a stellar wind or via RLOF.  In the latter case, the donors transfer 
all the material toward the accretor, whereas for the wind-fed systems only a fraction of the 
material is captured by the compact object.  For every accreting system the bolometric luminosity 
($L_{\rm bol}$) is calculated based on the secular average mass accretion rate (\.{M}$_{\rm acc}$) 
as 
\begin{equation}
L_{\rm bol} = \epsilon {G M_{\rm acc} \dot{M}_{\rm acc} \over R_{\rm acc}}
\end{equation}
where $G$ is gravitational constant, $M_{\rm acc}$ is the mass of the accretor, $R_{\rm acc}$ is the 
radius of the accretor (10 km for a NS and 3 Schwarzschild radii for a BH), and $\epsilon$ gives the 
conversion efficiency of gravitational binding energy to radiation associated with accretion onto a 
NS (surface accretion $\epsilon=1.0$) and onto a BH (disk accretion $\epsilon=0.5$). 

To determine the simulated X-ray luminosities of our population synthesis, we note that very 
little of the transferred material is accreted for transient systems during their inactive state. 
The spectra of neutron stars during this state consist of a soft blackbody component and a hard 
power law component with a tendency for a larger contribution from the power law component for 
sources characterized by lower quiescent luminosities (Jonker et al. 2004, see also Tomsick et al. 
2004).  The soft component is described by energies of $\sim 0.1 - 0.3$ keV (see Campana \& 
Stella 2003). The theoretical models based on deep crustal heating associated with pycnonuclear 
reactions (Brown, Bildsten, \& Rutledge 1998; Colpi et al. 2001) can provide an understanding 
of the black body-like nature of the spectrum and the observed emission level ($\sim 10^{32} - 
10^{33}$ ergs s$^{-1}$) without necessarily invoking the accretion of matter. However, short 
term variability during the quiescent state in Aql X-1 (Rutledge et al. 2002) and in Cen X-4 
(Campana et al. 2004) has also been observed, suggesting that accretion of matter may be 
occurring in Aql X-1 and Cen X-4 (Asai et al. 1996, 1998; Campana et al. 1998, 2000; Campana 2004).  
The accretion rate during this phase is likely to be at a low level since the inferred luminosities 
above 2 keV are $\sim 10^{31}$ erg s$^{-1}$. In contrast to the majority of transient systems in 
quiescence, the spectral energy distribution of the millisecond accreting X-ray pulsar, SAX 
J1808.4-3658, is dominated by a hard component with a luminosity of $\sim 5 \times 10^{31}$ ergs 
s$^{-1}$.  Such a nonthermal component may arise from synchrotron emission in a shock front 
produced by the interaction of a relativistic pulsar wind with matter from the companion 
star (Tavani \& Arons 1997; Campana et al. 1998).  Based on these observational results, we adopt 
$10^{31}$ ergs s$^{-1}$ as a lower limit for the hard X-ray luminosity, above 2 keV, for transient 
NS systems in quiescence, recognizing that the average luminosity level can be higher $\gtrsim 
10^{32}$ ergs s$^{-1}$ (e.g., Rutledge et al. 2001; Rutledge 2002; see Jonker et al. 2004). 
Due to the lack of a definitive theory for the hard X-ray emission component of transient 
systems in quiescent, we take a semi empirical approach for the quiescent NS transients and adopt an
X-ray luminosity level of $10^{31} - 10^{32}$ ergs s$^{-1}$ above 2 keV.  Furthermore, we assume that 
the quiescent NS transient X-ray luminosities are evenly distributed in the above range.

The quiescent emission from BH transient systems, on the other hand, is likely related 
to a low level of mass accretion.  In contrast to the NS transient systems, the spectra
are typically harder and are not described by a black body.  Observations of the 14 black hole 
transient systems for which the quiescent emission has been obtained reveal luminosities in the range 
from $\sim 10^{30} - 10^{33}$ ergs s$^{-1}$ with a median luminosity of $\sim 2 \times 10^{31}$ 
ergs s$^{-1}$ (see Tomsick et al. 2003). As for the NS transient systems a semi empirical approach
is adopted for the BH transient systems.  In particular, we assume that most (80\%) of the 
quiescent BH transient X-ray luminosities 
above 2 keV are evenly distributed in the $10^{30} - 10^{32}$ ergs s$^{-1}$ range, while the rest 
(20\%) of the systems are slightly brighter: luminosities evenly distributed in the 
$\sim 10^{32} - 10^{33}$ ergs s$^{-1}$ range  (see Fig. 3  of Tomsick et al. 2003).

Transient RLOF systems in outburst are bright X-ray sources, with luminosities
corresponding to a fraction ($\eta_{out}$) of the critical Eddington luminosity.  
The long period systems, with orbits that are sufficiently extensive for  
a large accretion disk to be formed, are usually found to emit at the Eddington 
luminosity ($L_{\rm edd}$) 
during outburst, while the outburst luminosities of short period systems are lower by 
about an order of magnitude. Therefore, we apply an X-ray luminosity correction factor at 
outburst corresponding to $\eta_{\rm out}=0.1$ and $\eta_{\rm out}=1$ for the short and long 
period systems respectively. The critical periods, over which the Eddington luminosity is 
adopted, are taken to be 1 day and 10 hrs for NS and BH transients in outburst respectively 
(Chen, Shrader \& Livio 1997; Garcia et al. 2003; T. Maccarone, private communication). 
 
For the persistent sources (both RLOF and wind-fed\footnote{Note that all
of the wind-fed systems are modeled here as persistent X-ray sources.}) and 
all transients in the 
outburst stage, where accretion is the dominant contributor to the observed 
luminosity, we apply a correction factor, $\eta_{\rm bol}$, to the bolometric 
luminosity to account for the fraction of energy emitted above 2 keV.  
For different types of systems one may expect corrections of the order of 
0.1-0.5.  However, for clarity and comparison with previous studies the same 
correction is applied to all the sources within the limits $\eta_{\rm bol}=0.1-1.0$, 
with $\eta_{\rm bol}=0.1$ chosen as a standard value.

Therefore, we can obtain the simulated X-ray luminosity ([${\rm ergs\ s}^{-1}$]) from:
\begin{equation}
 L_x = \left\{ \begin{array}{ll}
        10^{31} - 10^{32} & {\rm all\ quiescent\ NS\ transients} \\ 
        10^{30} - 10^{32} & {\rm 80\%\ quiescent\ BH\ transients} \\         
        10^{32} - 10^{33} & {\rm 20\%\ quiescent\ BH\ transients} \\
\eta_{\rm bol} \eta_{\rm out} L_{\rm Edd} & {\rm outburst\ NS/BH\ transients} \\
\eta_{\rm bol} L_{\rm bol} & {\rm persistent\ (RLOF\ and\ Wind\ fed)} \\
 \end{array} \right.
\end{equation}
where $L_{\rm Edd}$ represents the Eddington luminosity.

\subsection{Standard Model Calculations}

The simulated X-ray luminosity distributions of both wind-fed and RLOF-fed systems 
with NS and BH accretors are illustrated in Figure 1. 
The RLOF-fed systems dominate the overall X-ray population at 
luminosities greater than $L_x = 10^{30}$ ergs s$^{-1}$ with most of the systems 
lying within the range of $L_x \sim 10^{31} - 10^{32}$ ergs s$^{-1}$.  
The majority of these RLOF systems are transient X-ray sources in
quiescence formed through the AIC scenario presented in Belczynski \& Taam (2004).
The number 
of X-ray systems (435 of RLOF-fed and 56 of wind-fed sources) were calibrated to 
represent the Galactic center population in the MM03 survey and are also listed in 
Table 1 (for details see \S\,3.4).

In Table 1 the relative frequencies for various types of systems categorized by their 
donor and accretor found in our simulation are listed.  As expected from the examination 
of Figure 1, the RLOF-fed systems dominate ($88.6$\%) the population. The majority of 
these systems are NS transient sources in quiescence with typical luminosities of $L_x \sim 
10^{31} - 10^{32}$ ergs s$^{-1}$.  Most of the RLOF-fed systems have low-mass WD donors 
with the mass transfer rates of $\sim 10^{-11} \msun\ {\rm yr}^{-1}$ leading to transient 
behavior through a long lived ($\sim 1$\ Gyr) phase (for details see Belczynski \& Taam 2004). 
Over half of the systems belong to the ultra-short period LMXB class with orbital periods 
below 2 hrs (60.7\%), while the remaining RLOF-fed binaries (27.9\%) are characterized by 
longer periods extending beyond one month.

A small, but significant fraction of wind-fed systems ($11.4$\%) are found, 
mostly with low-mass ($M \leq 3 \msun$: 7.9\%) and intermediate-mass 
companions ($3 < M \leq 8 \msun$: 3.3\%). However, there is no statistically 
significant population of systems found with massive ($M > 8 \msun$)
companions (HMXBs). 
Most of the wind-fed systems are NS-MS binaries, however several have 
an evolved (red giant) donor, and a few have a BH accretor.
For the wind-fed systems, most are characterized by eccentric 
orbits with the most massive donors possibly exhibiting a B{\em e} 
X-ray binary phase, although we do not follow orbit-dependent accretion. 
 
Our X-ray population will remain confined to the Galactic center region since the spatial velocities 
induced as a result of the mass loss and nascent kick velocities accompanying NS/BH formation  are 
small compared to their mean spatial velocities in this region.  In particular, the average velocities 
acquired are $\sim 60-100$ km s$^{-1}$ for both the wind-fed and RLOF mass transferring systems. 
They are significantly smaller than the velocities of $\sim 200$ km s$^{-1}$ characteristic of the 
stellar motions in this region.  We point out that the velocities calculated for the RLOF mass 
transferring systems only refer to a minor subgroup (71 out of 435 systems, see Table 1) that have 
undergone a core collapse supernova since the majority of these systems are formed via the AIC 
mechanism with no nascent kick.  The low velocities are a direct consequence of the use of the 
recent NS kick distribution of Arzoumanian et al. (2002).

\subsection{Parameter study}

To determine the sensitivity of the results to the major uncertainties of the binary evolution, 
we have recalculated several models with different choices of evolutionary parameters.  
In particular, a model with a reduced efficiency of the CE ejection of $\alpha \times \lambda = 
0.1$  (in the standard model $\alpha \times \lambda = 1$) was simulated; for details see Belczynski 
et al. (2002).  In this case, the number of NS/BH X-ray binaries (XRBs) in the Galactic center is significantly 
decreased (109), with the relative number of wind-fed to RLOF-fed systems remaining comparable to 
the standard model.  The sensitivity of the model simulation to the assumed mass ratio distribution 
was also studied.  Specifically, a model in which both masses were drawn independently (as opposed 
to correlated initial masses for the two binary components via a flat mass ratio distribution in 
the standard model) was calculated.  This led to an increase in the number of systems which enter 
and merge in the CE phase, decreasing the number of NS/BH XRBs in the Galactic center (296) and 
lowering the importance of RLOF-fed systems relative to wind-fed systems (245 RLOF-fed 
systems compared to 51 wind-fed systems).  This decrease in the number of XRBs for the above 
two models provides an indication of the influence of the uncertainties associated with our 
treatment of the CE evolution phase (for details see Belczynski \& Taam 2004). 

Since the stellar wind mass loss rates are somewhat uncertain, an additional model with the stellar 
wind mass loss rates reduced by a factor of 2 for all mass losing stars was calculated. The smaller 
wind mass loss rates have two opposing tendencies on wind-fed systems.  First, there is an 
obvious decrease of the mass accretion rate, and thus a decrease in the number of wind-fed X-ray 
systems above certain critical value of $L_x$.  Second, the systems tend to be tighter, due to the 
fact that less mass is carried away (Jeans mode of mass loss).  
Therefore, the fraction of matter lost in the wind that is captured increass, leading to a tendency 
for an increase in the accretion rate.  The latter effect dominates over the former, and there is 
an overall increase in the number of wind-fed systems (82) as compared to the standard model (56).
The number of RLOF-fed systems is not greatly affected (416).

A model in which we use the full natal kicks (Arzoumanian et al. 2002) for NSs formed via AIC as 
opposed to no AIC kicks in the standard model was also calculated.  In this model wind-fed systems 
are not affected, however a number of RLOF systems progenitors are disrupted at the formation of 
the NS. Additionally, many of the surviving RLOF systems formed in AIC acquire high ($ > 200 {\rm km}\ 
{\rm s}^{-1}$) systemic speeds, exceeding the escape velocity from the Galactic center (but still 
bound to the Galaxy). This follows 
from the result that the AIC channel occurs, in general, for systems at short orbital periods (where 
the components orbit about their common center at high velocities), and the systems which 
survive can acquire a significant additional systemic speed (comparable to the component's orbital 
velocity).  The total number of NS/BH XRBs decreases to 201, while the relative contribution of 
wind-fed sources in the X-ray population significantly increases (see Fig.~3) as compared to the 
standard model. 

An additional model, without tidal binary component interactions, was calculated to compare the 
results to the other studies.  In this model, we find the highest number of wind-fed sources (87) 
in the simulations and a slight decrease of RLOF-fed systems (387) in comparison with our standard 
model calculation.  In the absence of tidal interaction, the orbital expansion associated with 
transfer of spin angular momentum to the orbit does not take place and the X-ray binary progenitors 
are generally found in shorter period orbits.  For the tightest systems (progenitors of RLOF sources) 
this leads to more frequent mergers in CE episodes (decreasing the formation rates).  On the other 
hand, the absence of tidal effects leads to tighter orbits for wider systems (the progenitors of 
wind-fed sources) and, therefore, to brighter systems resulting from higher wind capture rates,  
increasing their number in the studied sample. In particular, the X-ray luminosities 
of wind-fed sources can exceed $L_x \sim 10^{34}$ ergs s$^{-1}$ for
intermediate mass X-ray binaries. 
In this calculation, similar to the standard model, the number of HMXBs is
negligible, as expected for the small sampled volume corresponding to 
the MM03 survey. 

We also relaxed our assumption on the bolometric correction, and recalculated the NS/BH system 
X-ray luminosities with $\eta_{\rm bol}=1.0$ (no bolometric correction). As expected (see eq. 2), 
only the brightest sources in RLOF class (either persistent or transients in outburst) are affected, 
with $L_x$ extending to $\sim 10^{38}$ ergs s$^{-1}$.  Since the wind-fed systems are brighter 
(by the factor of ten), the number of wind-fed sources increases to 80 as additional systems 
are brighter than our adopted critical threshold luminosity ($L_x = 10^{30}$ ergs s$^{-1}$).

Throughout our study we have allowed for the possibility of accretion induced
collapse of a heavy WD to a NS. The specific arguments and new input physics
supporting AIC was presented and discussed in detail by Belczynski \& Taam (2004). 
As an indication of its importance, we examine the implications of its neglect 
on the results of our study. The primary effect is a decrease in the number of 
NS/BH X-ray sources to 126 (as compared with 491 in standard model, see also Table 1).  
This decrease results in a 
significant depletion of RLOF population (71), while wind-fed
sources would remain virtually unchanged (55). In this model the majority of RLOF 
systems are found with long periods (45), but some systems are still found within
ultrashort period class (26). As a result, the 
number of transients and the number of bright sources decrease. 
The shape of the X-ray luminosity distribution is not significantly 
altered, although the relative contribution of wind-fed to RLOF-fed sources is increased. 

Notwithstanding the models with rather extreme assumptions on accretion induced 
collapse (full natal kicks applied during NS formation in the AIC process
and a model in which AIC is not allowed) the RLOF-fed NS/BH XRBs 
are found to be more numerous than their wind-fed counterparts by more than a 
factor of 2. For comparison, the X-ray luminosity distributions for the most extreme model 
assumptions are shown in Figures~2 and 3. Note the change of vertical scale
between different plots. For all of the models most systems are found at low
luminosities ($L_x \sim 10^{30}-10^{33}$ ergs s$^{-1}$), with several bright systems 
usually found at ($L_x \sim 10^{36}-10^{38}$ ergs s$^{-1}$). 
Although the shape of the distribution is very similar, the number of sources 
and the ratio of wind-fed sources to RLOF sources may change significantly from model
to model.  

\subsection{Expected numbers}

The total mass contained in stars corresponding to our simulation with $10^6$ massive binaries 
(primary component more massive than $4 M_\odot$) can be estimated based on the three component 
broken law of Kroupa et al. (1999). By extending our simulation mass range to $0.08 M_\odot$ and 
including a 50\% contribution due to single stars, we estimate the total mass to be $\sim 1.5 
\times 10^8 M_\odot$. As based on the recent mass model by Launhardt, Zylka, \& Mezger (2002), 
this is comparable to the stellar mass in a cylindrical radius of 20 pc and a depth  of 440 
pc centered on the Galactic center (Muno et al. 2004). We note, however, that the stellar mass 
would be overestimated by a factor of 5 if the MM03 survey corresponded to a sphere of radius 
20 pc.  Our results (Table 1, also previous subsections) can also be applied to the  
WGL02 survey by taking into account that the larger field is about 4 times more massive than 
modeled in our population synthesis.  
In our standard simulation it is found that there are $\sim 500$ and $\sim 2000$ NS/BH X-ray 
binaries with luminosities $L_x > 10^{30}$ ergs s$^{-1}$ (see Table 1) for the MM03 and WGL02 fields, 
respectively. However, the majority of the synthetic NS/BH accretion sources are expected to be 
characterized by low X-ray luminosities ($L_x \sim 10^{31}-10^{32}$ ergs s$^{-1}$) in the hard X-ray 
band (see Figure 1, 2 and 3).  We note that many of our simulated sources can have a significant soft 
component with higher luminosities  (see discussion in the \S\,3.1). 

The majority ($\sim 500$) of the Galactic center sources discovered in the WGL02 survey were 
characterized  
by $L_x \sim 10^{33}-10^{35}$ ergs s$^{-1}$ and hard spectra.  In our standard model calculation, 
there are no statistically significant sources in that luminosity range.  However, 
in the model with no bolometric correction applied ($\eta_{\rm bol}=1.0$), 44 
sources in that range are found, all of which are wind-fed systems. This 
simply reflects the fact, that all of the wind-fed systems found at $L_x \sim 10^{32}-10^{33}$ 
ergs s$^{-1}$ in the standard model, are shifted to the higher luminosities with $\eta_{\rm bol}=1.0$.
In addition, for the models with altered assumptions on AIC (either full AIC kicks or AIC not allowed) 
the number of wind-fed systems (amounting to the same as for the standard model) dominate over 
RLOF-fed binaries in this specific luminosity range. 
Although these systems are promising candidates for the unidentified X-ray population, it is 
likely that they represent a small fraction ($\lesssim 20$\%) of the observed sources in WGL02 
survey. Our results suggest that other sources, and not the accretion powered NS/BH binaries 
constitute the majority of point sources in the WGL02 survey.

In the deeper survey of MM03 a faint source population of 2079 point sources, characterized by $L_x \sim 
10^{30}- 10^{33}$ ergs s$^{-1}$ was discovered. The spectra of these sources were fit to an 
absorbed power law with photon index, $\Gamma$.  The majority of these sources is described by 
hard spectra $\Gamma < 1$ (1427), with an important subpopulation characterized by
softer spectra $\Gamma > 1$ (652\footnote{numbers from M.~Muno, foreground
sources excluded; private communication}).  We note that the majority 
of our synthetic XRBs have luminosities in a range very similar to these found 
for point sources in MM03 with a significant number of these systems corresponding to 
ultrashort period RLOF-fed NS-WD binaries. Due to the very low mass transfer rates in these 
binaries, the sources are expected to be transient and would most 
likely be observed during their quiescent state at low luminosities.  Muno et al. (2004) suggest that 
the faint hard source population ($\Gamma < 1$) can be consistent with magnetic cataclysmic variable 
systems. It is suggested that the transient RLOF-fed NS-WD transients in quiescence 
studied in this paper may contribute to the faint population characterized 
by softer spectra. 

For the bright systems, 15 XRBs are found in our simulated population of 
the MM03 field. These sources are characterized by luminosities 
$\sim 10^{36}-10^{38}$ ergs s$^{-1}$ and are RLOF persistent and RLOF 
transient systems in outburst. However, only a few bright transients have been 
observed in the field of the MM03 survey (M.Muno, private communication), 
suggesting that the simulation leads to an overproduction of such systems 
(see \S 4).

For the WGL02 field, our standard model calculation shows $\sim 60$ X-ray sources 
brighter than $10^{36}$ ergs s$^{-1}$. Most of these systems in our
simulation are persistent X-ray sources.
However, only $\lesssim 20$\ bright sources were observed (WGL02) with most as 
transient systems in outburst in the Galactic Center. 
This suggests that the parameters intrinsic to our standard model
affecting the bright population require modification. For the alternative models 
(full AIC kicks, lowered CE efficiency, reduced stellar winds and AIC not allowed) it is 
found that number of the bright sources is reduced to less than $\sim 20$. We note 
that the number of transient sources exceeds the number of persistent sources only 
for the simulation with reduced stellar winds. 

To account for the entire bright Galactic field XRB population we increase the above number 
by factor of $\sim 100$, yielding $\sim 6000$ bright systems (4000 persistent 
and 2000 transients in outburst).  This is an overestimate, since 
only 150 of these systems are observed.  This overproduction of bright systems is common to all 
our models (including the neglect of AIC) and the other 
studies of the RLOF X-ray sources in the Galaxy (e.g., Pfahl, Rappaport, \& Podsiadlowski 2003)
and implies an overproduction of the bright population in the Galactic center as well. We note 
that the overproduction may not be as great for the faint population since the relative number 
of faint to bright sources can change from model to model. In particular the number of bright 
sources decreases by a factor of 5, whereas the number of faint sources is relatively unchanged 
for a model with reduced stellar winds. 

\subsection{Comparison with other studies}

Willems \& Kolb (2003) estimated the number of pre-LMXBs with main sequence companions less 
than about $2 \msun$ in the Galaxy. Their number may be directly rescaled to the smaller 
expected population in the Galactic center, yielding $\sim 10^2-10^3$ sources for the WGL02 
survey field (B.Willems, private communication).  Since the study of Willems \& Kolb (2003) 
assumes no bolometric correction, we use our model with no bolometric correction for comparison. 
In this case we expect $\sim 250$ wind-fed low-mass X-ray binaries (pre-LMXBs), a number which 
is consistent with the Willems \& Kolb (2003) findings.  If the bolometric correction was 
applied, the expected number of pre-LMXBs with luminosities $\gtrsim 10^{30}$ erg s$^{-1}$ would 
decrease by a factor of $\sim 2$. 

A direct comparison with PRP02 is not easily accomplished since the initial input (e.g., IMF, 
eccentricities) and treatment of the RLOF phases as well as model assumptions (e.g., natal 
kicks) differ.  Despite the model differences, it is found that the numbers of wind-fed 
systems predicted by PRP02 and in the present study for the WGL02 survey can be comparable 
(see below).  In addition the X-ray luminosity distributions show that the majority of the 
synthetic wind-fed sources of both studies are characterized by rather low-luminosity $L_x \sim 
10^{32}-10^{33}$ ergs s$^{-1}$ (although we are unclear about the adopted choice for the 
bolometric correction in PRP02). The primary difference stems from the fact that the wind-fed 
sources in our study are characterized by low- and intermediate-mass MS companions, while PRP02
accounted only for the wind-fed sources with either intermediate- or high-mass companions. 

In the present study the intermediate mass X-ray binaries (IMXBs) greatly dominate over 
HMXBs, in contrast to that found by PRP02. In fact, PRP02 obtain in their model K2 about 
equal numbers of HMXBs and IMXBs. This arises from their specific choice of kick scenario, 
which favors the survival of HMXBs.  We prefer to use the observationally derived distribution 
of kicks (Arzoumanian et al. 2002).  With the use of rather standard kick velocity distribution 
(model K1), the PRP02 calculation results in the dominance of IMXBs over HMXBs (by a factor of 
about 5-10; as one can infer from the middle panel of their Figure 1). This result is already 
comparable to the results obtained in the present study. Moreover, there is an additional 
factor which lowers the number of HMXBs as compared to IMXBs within our population synthesis 
model.  Specifically, PRP02 use a constant escape wind speed independent of mass (or spectral
type). However, it is known that this speed depends on a mass of star (e.g.  Lamers, Snow \& 
Lindholm 1995). For the mass range used in our study it varies by about an order of magnitude. 
The capture rate (and thus resulting X-ray luminosity of wind fed system) depends strongly on this 
velocity.  The higher the wind velocity, the smaller the capture rate and the lower the X-ray 
luminosity.  In our simulation the most massive stars have wind speeds as high as 5 times the escape speed 
(Lamers et al. 1995), while PRP02 considers wind speeds not greater than twice the escape speed. 
This effect results in a smaller number of bright X-ray binaries with massive companions (HMXBs) as
compared to PRP02 study.

For the WGL02 survey field PRP02 obtain a total of 250 and 600 wind-fed neutron star systems with 
main sequence companions more massive than $3 \msun$ for their kick models K1 and K2. 
The number of PRP02 systems is reduced due to the detection limit appropriate for WGL02 survey 
($L_x=10^{33}$ ergs s$^{-1}$) by a factor of 2.5 for a wind speed equal to the escape velocity or 
by a factor of 20 for a wind speed equal to twice the escape velocity.  This leads to a reduction in the 
number of systems in the PRP02 K1 model from 250 to 12-100 systems in the WGL02 field.

Since the K1 kick distribution of PRP02 (a single Maxwellian with $\sigma=300$ km s$^{_1}$) 
is comparable to the present study (bimodal distribution with 40\% kicks with $\sigma=90$ km 
s$^{-1}$ and 60\% with 
$\sigma=500$ km s$^{-1}$ - Arzoumanian et al.  2002), we compare our results with the results from 
model K1. Our total number of wind-fed sources (standard model) for WGL02 field is 224 (see Table 1; 
note an increase by factor of 4 for the WGL02 field). In the following we give the different 
corrections factors in order to compare our results to that obtained by PRP02.  First in the model 
with no bolometric correction (assumed to be the case in PRP02) we expect a greater number of wind-fed 
sources (320).  Of these, only a 44 have luminosities in excess of $10^{33}$ ergs s$^{-1}$. Those 
systems with MS companions which are more massive than $3 \msun$ amounts to 32.  Therefore, we 
conclude that the overall number of wind fed neutron star systems systems in PRP02 study for model 
K1 is in approximate agreement with the results obtained in our study for similar model assumptions.
However, PRP02 increase the number of WNS by using their alternative kick model (K2), which can 
reproduce the entire population of X-ray sources in the WGL02 field. 

\section{Discussion}

The population of low luminosity X-ray binaries in the Galactic center region has 
been investigated within the framework of a population synthesis technique.  Although 
it has been found that this population may be more heterogeneous than previously predicted, 
our results suggest that accreting neutron star systems are not likely to be the major 
contributor to the faint X-ray source population in the Galactic center.  

The RLOF systems investigated in this paper provide an additional low luminosity 
component.  Many of these additional systems 
are composed of neutron stars with white dwarf companions in ultra-short orbital periods 
($\lesssim 2$\ hrs) in which mass is transferred at rates ranging from  
$10^{-12}-10^{-10} \mpy$ as a result of the action of angular momentum losses associated
with gravitational radiation.  Because the mass transfer rates are low, these 
systems are expected to exhibit transient quiescent/outburst behavior.  For a small duty cycle 
of $\sim$\ 1\% it is expected that most of these sources would be found in their quiescent 
state with X-ray luminosities in the range of $\sim 10^{30}-10^{33}$ erg sec$^{-1}$.

The population of hard X-ray sources ($10^{33}-10^{35}$ erg sec$^{-1}$) discovered in 
WGL02 survey was interpreted in the context of wind-fed NS-MS binary systems (PRP02). 
We find that such systems can be a contributor to the X-ray population in this 
luminosity range provided that $\eta_{bol} \sim 1$.  However, our results cannot explain 
the entire population since the predicted number of such systems ($\sim 30$) is
much lower than the number of observed sources ($\gtrsim 100$). 

The deeper MM03 exposure revealed a large population of faint ($L_x \sim 10^{30}-10^{33}$ erg 
sec$^{-1}$) point sources. The majority of these sources are characterized by hard spectra, 
and are too numerous to be explained by wind-fed NS/BH systems and too hard to be explained by 
RLOF-fed NS/BH systems.  Muno et al. 2004 have suggested that the intermediate polar class 
systems can explain both their spectral properties and their observed number.  The nature of 
the population characterized by softer spectra is unknown, although a number of candidates 
have been proposed.  We have found that the number of RLOF-fed NS/BH transient systems in 
quiescence could be as high as 400 in comparison to the 652 observed, but this leads to an 
overproduction of bright LMXBs in the Galaxy.  The sensitivity of our results to input 
parameters have been explored, which can lead to a reduced contribution of RLOF sources and 
to an overall reduction in the number of synthetic soft NS/BH X-ray sources to a level further 
below that observed in MM03 survey. 

The identification of the faint soft X-ray sources will be essential for confirming their nature. 
Such studies may be fruitfully carried out in the low extinction regions near the Galactic center
where the contribution from the low energy component of their spectra (corresponding to less 
than 1 keV) may be detectable. At these lower energy ranges, the X-ray luminosities of our 
proposed candidates could be as high as $L_x \sim 10^{32}-10^{33}$ ergs s$^{-1}$. The Chandra 
and HST surveys of Baade's window have been carried out (Grindlay et al. 2003), and the forthcoming 
results may have bearing on the importance of this population.

\acknowledgements
We would like to thank M.Muno, D.Wang, T.Maccarone, B.Willems, P.Podsiadlowski and N.Ivanova 
for useful discussions on the project and the anonymous referee for the comments 
which have helped to improve our study. 
This research was supported in part by the NSF under Grant No. AST-0200876 to 
RT and by the KBN Grant PBZ-KBN-054/P03/2001 to KB.

\pagebreak

\psfig{file=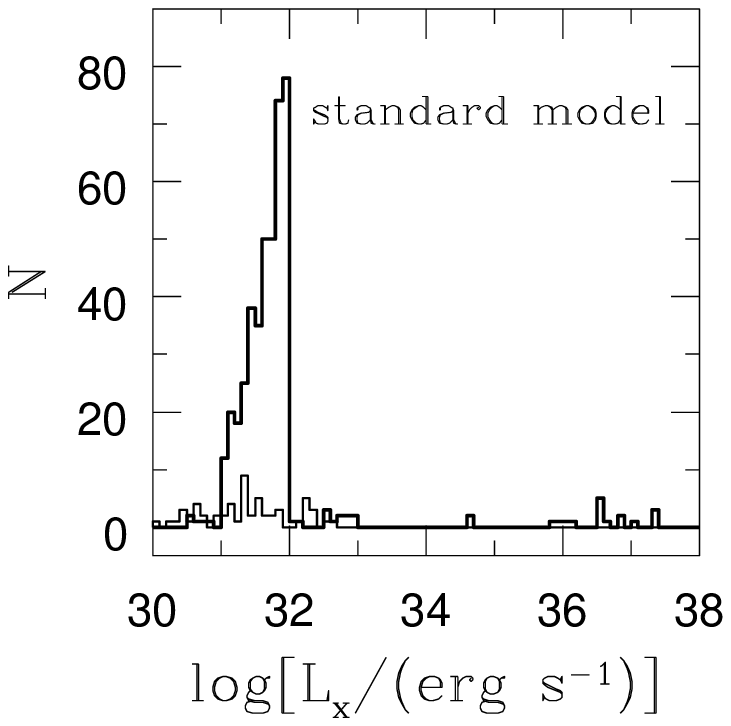,width=0.45\textwidth}
\figcaption[]{
\footnotesize
X-ray luminosity distribution for RLOF-fed (thick line) and wind-fed 
(thin line) binaries for the standard model calculation described in
(\S\,3.2).  
Note the very clear dominance of RLOF-fed systems in the faint luminosity 
regime ($L_x \sim 10^{31} - 10^{32}$ erg sec$^{-1}$) for our simulated 
observed sample of Galactic center X-ray binaries.
} 

\pagebreak

\psfig{file=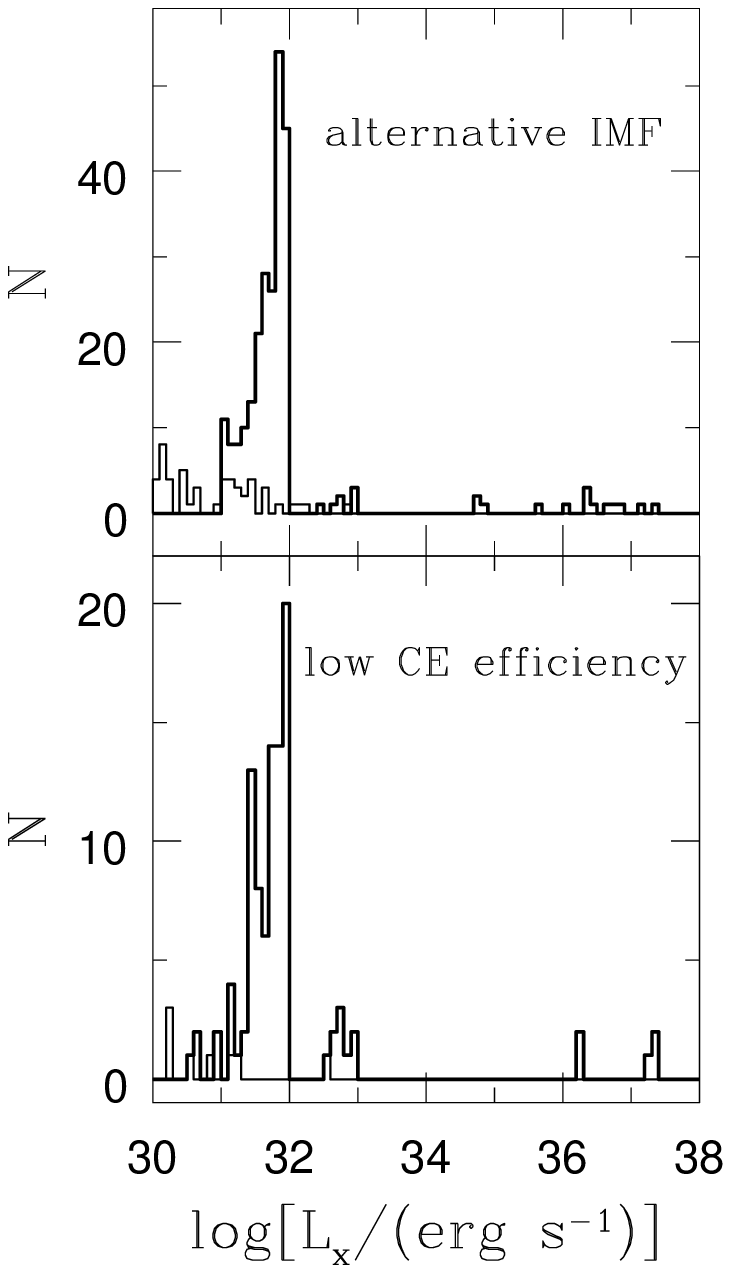,width=0.45\textwidth}
\figcaption[]{
\footnotesize
X-ray luminosity distributions for RLOF-fed (thick line) and wind-fed 
(thin line) binaries for different assumptions on the initial mass
distribution of binary components (top panel) and common envelope efficiency 
as compared to our standard model. For details of alternative model 
assumptions see text (\S\,3.3). Note the change in vertical scale on the 
panels.
} 

\pagebreak

\psfig{file=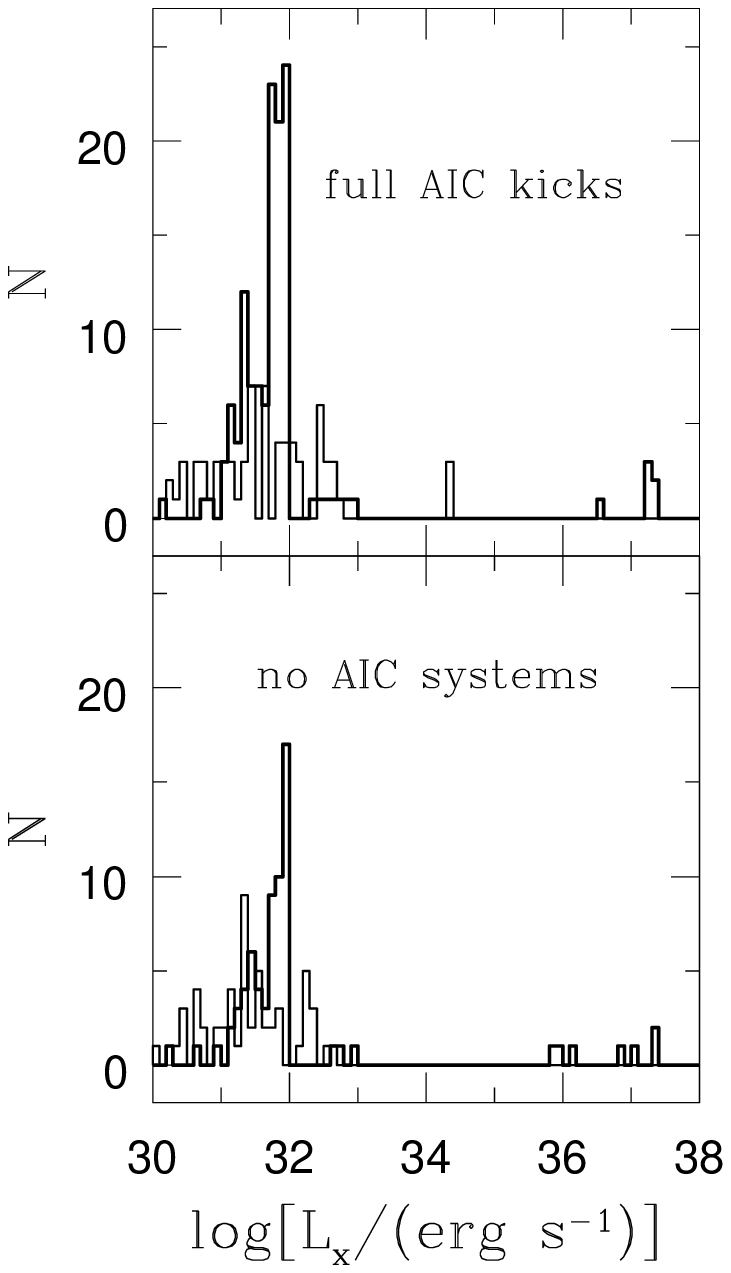,width=0.45\textwidth}
\figcaption[]{
\footnotesize
X-ray luminosity distributions for RLOF-fed (thick line) and wind-fed 
(thin line) binaries for different assumptions concerning accretion induced
collapse: full natal kicks incorporated (top panel) and AIC systems removed
from the sample. Details on the models in \S\,3.3.
} 

\pagebreak

\begin{deluxetable}{crr}
\tablewidth{250pt}
\tablecaption{ X-ray Binaries in the Galactic Center}
\tablehead{ Type\tablenotemark{a}& Wind-fed & RLOF-fed}
\startdata

Total:                      &  11.4\% (56)	 &   88.6\% (435) \\
NS accretor                 &  11.0\% (54)       &   74.1\% (364) \\
BH accretor                 &    0.4\% (2)	 &   14.4\% (71)  \\
                            &                    &                \\ 
MS donor                    &  10.2\% (50)	 &   21.6\% (106) \\
giant donor                 &    1.2\% (6)       &     7.7\% (38) \\
WD donor                    &      0\% (0)       &   59.3\% (291) \\
           	            &             	 &                \\
transient\tablenotemark{b}  &      0\% (0)       &   85.9\% (422) \\
ultrashort\tablenotemark{c} &      0\% (0)       &   60.7\% (298) \\
SN system\tablenotemark{d}  &  11.2\% (55)       &    14.4\% (71) \\

\enddata
\label{numbers}
\tablenotetext{a}{
X-ray active binaries ($L_x > 10^{30}$ erg sec$^{-1}$) are listed.
The relative occurrence frequency is given, followed by the occurrence 
frequency of specific types of accretors and donors 
(MS--main sequence star, WD--white dwarf, NS-neutron star, 
BH-- black hole). 
In parenthesis we list the expected number of sources in the Galactic 
center region corresponding to the Muno et al. (2003) survey (see 
\S\,3.4 for details).}
\tablenotetext{b}{All the transients are listed here. For the assumed duty
cycle of 1\%, only 3--5 of these systems are expected to be observed in the
outburst state.}
\tablenotetext{c}{Systems with orbital periods shorter than 2 hrs.}
\tablenotetext{d}{Systems in which there was a SN explosion.} 
\end{deluxetable}

\end{document}